\documentclass[referee]{aa}
\usepackage{graphicx}

\def\url#1{{\bf #1}}

\def\m@th{\mathsurround=0pt}
\def\EQM#1{\vcenter{\normalbaselines\m@th
    \ialign{${\displaystyle ##}$\hfil&&\ ${\displaystyle ##}$\hfil\crcr
    \mathstrut\crcr\noalign{\kern-\baselineskip}
    \noalign{\smallskip}
    #1\crcr\mathstrut\crcr\noalign{\kern-\baselineskip}}}}

\def\apjl{ApJ}%
\def\apjs{ApJS}%
\def\ao{Appl.~Opt.}%
\def\aap{A\&A}%
\def\prd{Phys.~Rev.~D}%
\def\ijmpD{Int.~Jour.~Mod.~Phys.~D}
\def\planss{Planet.~Space~Sci.}%
\begin{document}

\title{Gravity tests with INPOP planetary ephemerides.}

\author{A. Fienga\inst{1,2}
\and J. Laskar\inst{1}
\and P. Kuchynka\inst{1}
\and C. Leponcin-Lafitte\inst{4}
\and H. Manche\inst{1}
\and M. Gastineau\inst{1}
}

\institute{Astronomie et Syst\`emes Dynamiques,
IMCCE-CNRS UMR8028,
77 Av. Denfert-Rochereau, 75014 Paris, France
\and
Observatoire de Besan\c con- CNRS UMR6213, 
41bis Av. de l'Observatoire, 25000 Besan\c con
\and
ESAC
Darmstad, Germany
\and 
SYRTE UMR, 
Observatoire de Paris,
77 Av. Denfert-Rochereau, 75014 Paris, France
}

\offprints{A. Fienga, agnes.fienga@obs-besancon.fr}

\date{\today}

\titlerunning{Gravity tests with INPOP}
\authorrunning{Fienga et al}


%
\abstract{In this paper, we present several gravity tests made in using the last INPOP08 planetary ephemerides. We first propose two methods to estimate the PPN parameter $\beta$ and its correlated value, the Sun J2 and we discuss the correlation between the Sun J2 and the mass of the asteroid ring. We estimate possible advance in the planet perihelia. In the end we show that no constant acceleration larger than 1/4 the Pioneer anomaly can affect the planets of our solar system.
}

\maketitle

\section{Introduction}
\label{intro}

Thanks to the high precision achieved with the observations deduced from the tracking of spacecrafts, it becomes possible to estimate relativistic parameters, for instance mainly $\gamma$ and $\beta$, of the Parametrized Post-Newtonian formalism of General Relativity (Will, 1993). 
Nevertheless if $\gamma$ plays a role in the equations of motion, it is worth to note that light propagation is only sensible to that parameter. PPN $\gamma$ can be then estimated with high accuracy by light deflection measurements by VLBI (Shapiro et al. 2004), by time delay during an interplanetary roundtrip and by Doppler tracking data of a space mission (see for instance the Cassini experiment, Bertotti et al. 2003). 
This is also why, in the following, we put $\gamma=1$ in order to only test the sensitivity of PPN $\beta$ on the perihelion's advance of planets.
 However a physical quantity relative to the Sun, its oblateness $J_2$, plays also a key role in this phenomena. Indeed, usual expression of the advance of perihelion is given by (Will 2006)
\begin{equation}
\Delta\omega=\frac{2\pi(2\gamma-\beta+2) GM_{sun}}{a(1-e^2)c^2}+\frac{3\pi J_2 R^2_{sun}}{a^2(1-e^2)^2}\,
\label{omega}
\end{equation}
where $G$ and $c$ are the newtonian gravitational constant and the speed of light in a vacuum, respectively; $J_2$ and $R_{sun}$ are the Sun oblateness and the Sun equatorial radius; $a$ and $e$ are the semi-major axis and the eccentricity of the precessing planet. One can immediately suspect that it is not possible to do a relevant estimation of PPN $\beta$ without considering Sun $J_2$. 
Furthermore, it may be not possible to decorrelate safely these two quantities with only one planet. With INPOP08 (Fienga et al. 2009), MEX and VEX tracking data have lead to an important improvement of Mars and Venus orbits, respectively. It is then suitable to take advantage of this new situation by attempting to decorrelate these parameters. 
 
Indeed, the impact of the Mars and VEX observations is not only limited to the improvement of the planet dynamics. 
They also play a role in the determination of parameters such as the asteroid masses, the oblateness of the Sun and the PPN parameter $\beta$.  The ratio between the uncertainties of the observations and the sensitivity of the observed orbit to the GR modifications has been evaluated by dividing the cumulative advance of the perihelion over a period of time corresponding to the time span of observations by the angle uncertainty of INPOP and presented in table \ref{GRJ2}. If the amplitude of the advance of the perihelion on Venus and Mars orbits is considered for a set of observations of equivalent accuracy, Venus data will be seven times more efficient to test general relativity and to estimate the sun J2 than Mars.
 If VEX mission is prolongated from 2 years to 4 years and if VLBI  observations are done from the tracking of the spacecraft with an accuracy of about 1 mas, VEX data will be then as important for the PPN testing and Sun J2 estimations as the direct 800-meter accuracy radar ranging on Mercury.
Besides, the Mars data are still very important because of the long time span of observations of very good quality obtained since the Viking mission in 1978.

Thanks to the informations brought by the combination of very accurate tracking data of spacecraft orbiting different planets, the planetary ephemerides become then an interesting tool for gravity testing. In the following, we give some examples of such tests.

\begin{table}
 \caption{The first 2 columns give the a-priori INPOP uncertainties in geocentric angles and distances limited by the observation accuracies. In the third column, one may find the estimation of the general relativity and sun oblateness effect of the advance in perihelion, $\dot{\omega}$, on the Mercury, Venus and Mars perihelia per year. The fourth column gives the S/N ratio estimated over the period of time given in years in column 5.}
 \begin{tabular}{l c c r r l | l c c r r l}
 \hline
         &    INPOP        &   accuracy &  $\dot{\omega}$    & S/N &    period &         &    INPOP        &   accuracy &  $\dot{\omega}$    & S/N & period\\
 Planets    &        angle    &distance&    ''/yr    & & years    &  Planets    &        angle    &distance&    ''/yr    & & year\\
 \hline
 
 Venus    &        0.001''    & 4m&    0.086     &172&2 & Mars    &        0.001''    & 2m    &0.013     & 130    & 10 \\
         &                &    &      & 344    & 4&         &            &     &    & 390    & 30\\
 Mercury    &        0.050''    &1km&    0.43     & 300    & 35& & & & & &\\
 \hline
 \end{tabular}
 \label{GRJ2}
 \end{table}

\section{Determination of PPN $\beta$ and sun oblateness J2}
\label{betaJ2}

\subsection{Correlation between sun J2 and asteroid modeling}

The advance of the perihelion induced by general relativity and sun J2 has an impact very similar to the advance induced by the main-belt asteroids on inner planet orbit. 
In INPOP08, a ring was fixed to average the perturbations induced by the main-belt asteroids which cannot have their signal fitted individually on tracking observations. This ring has its physical characteristics (mass and distance to the sun) estimated independantely from the fit by considering the albedos and physical properties of 24635 asteroids (for more details see for instance Kuchynka et al. 2008).

\begin{figure*}
\begin{center}

\includegraphics[width=6cm]{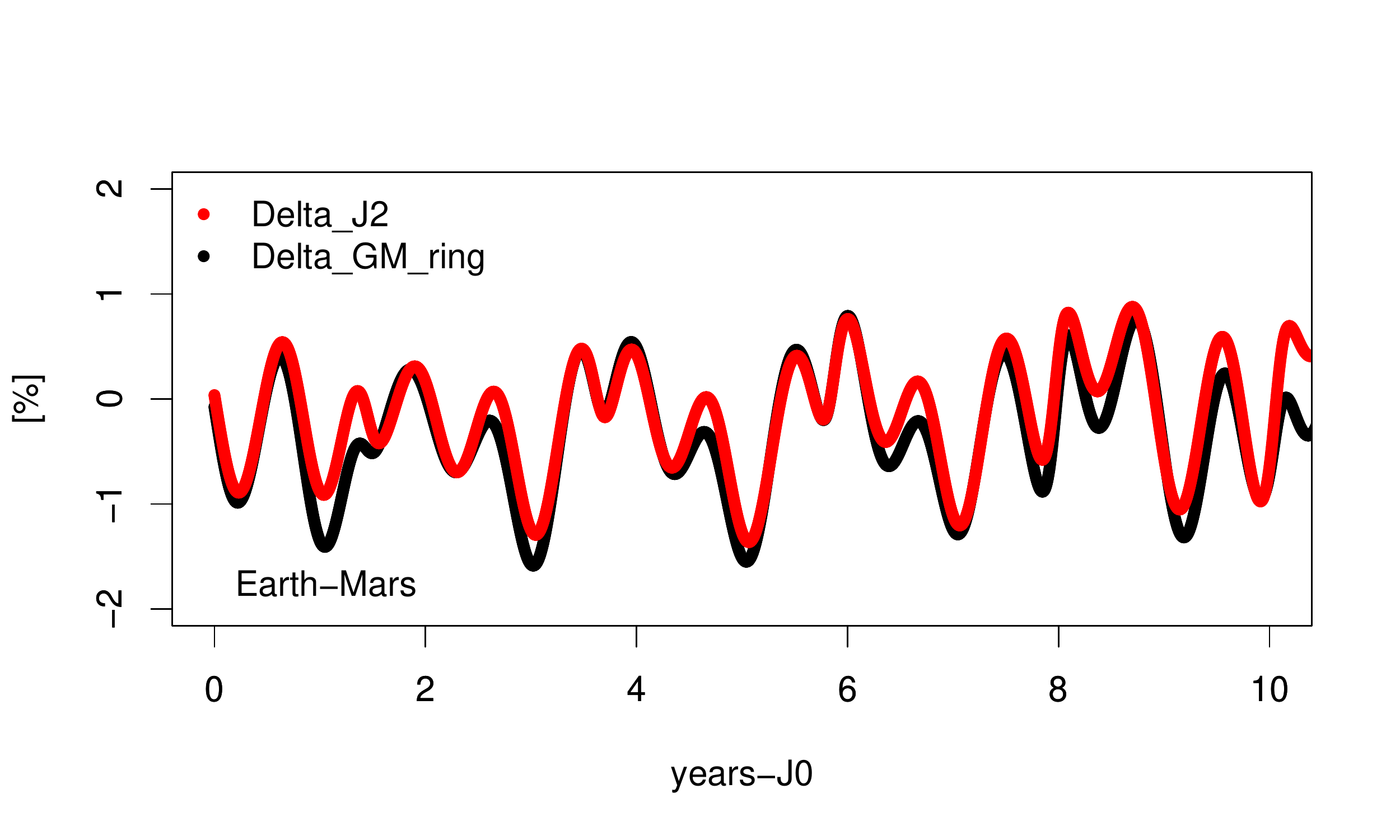} \includegraphics[width=6cm]{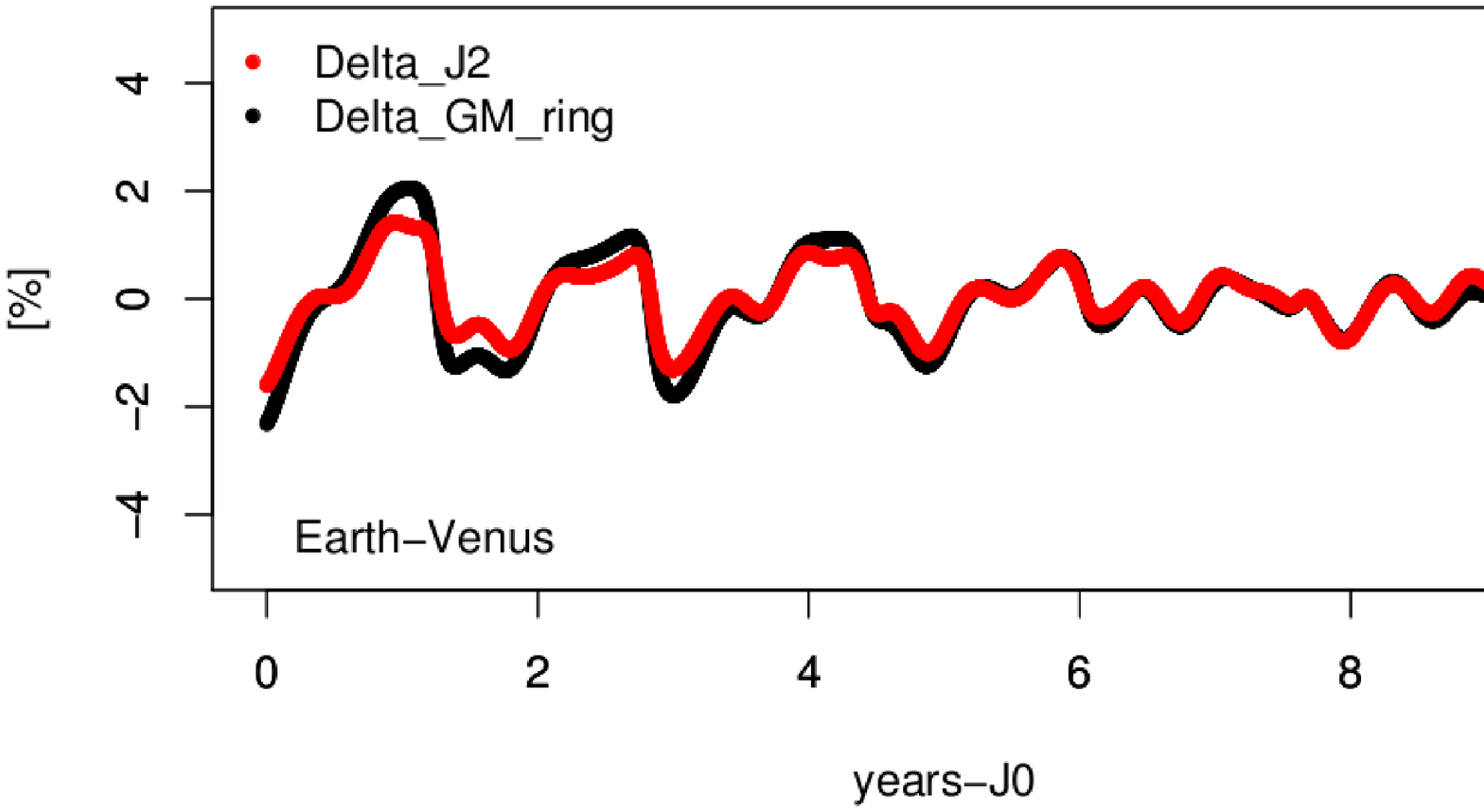}
 \includegraphics[width=6cm]{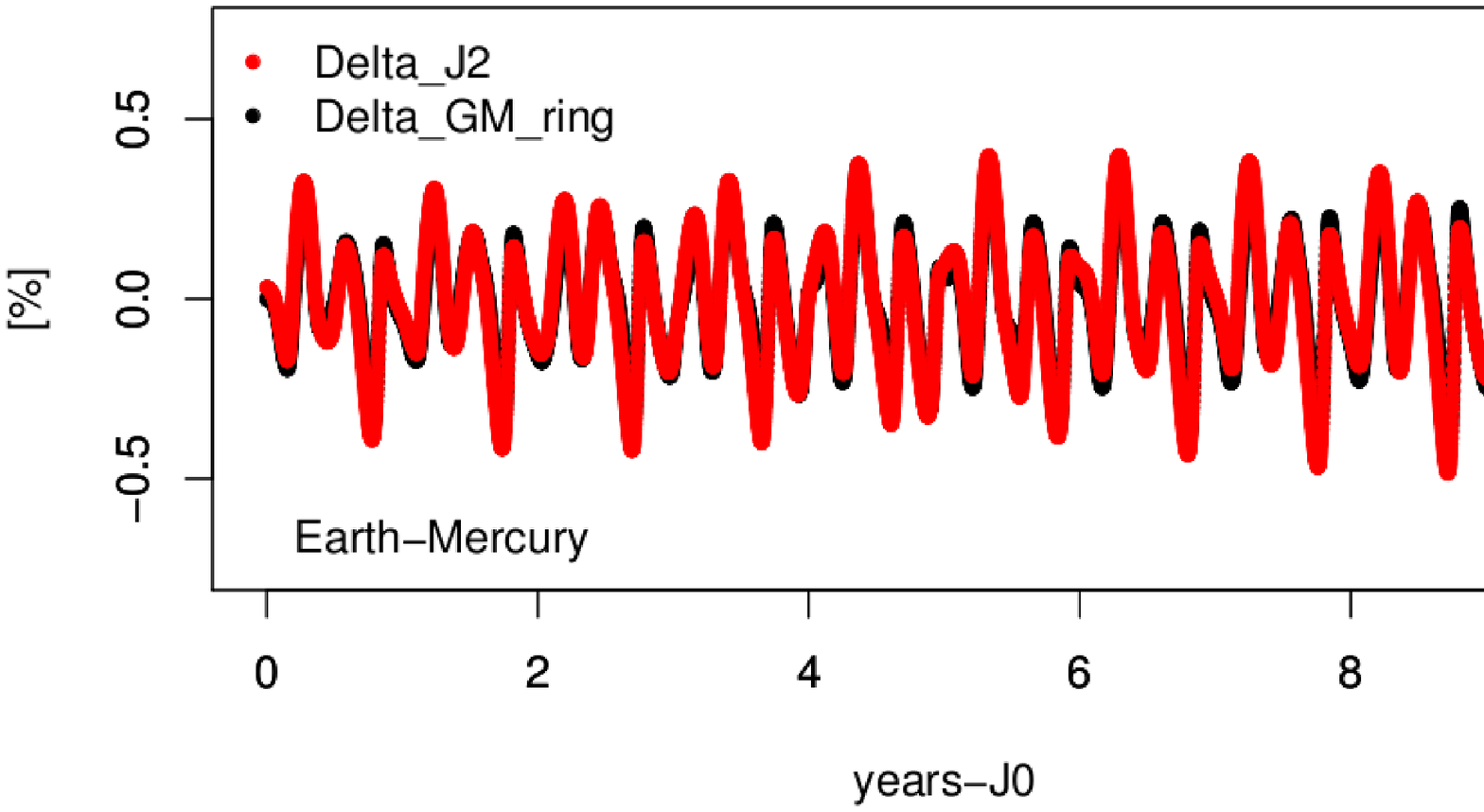} \includegraphics[width=6cm]{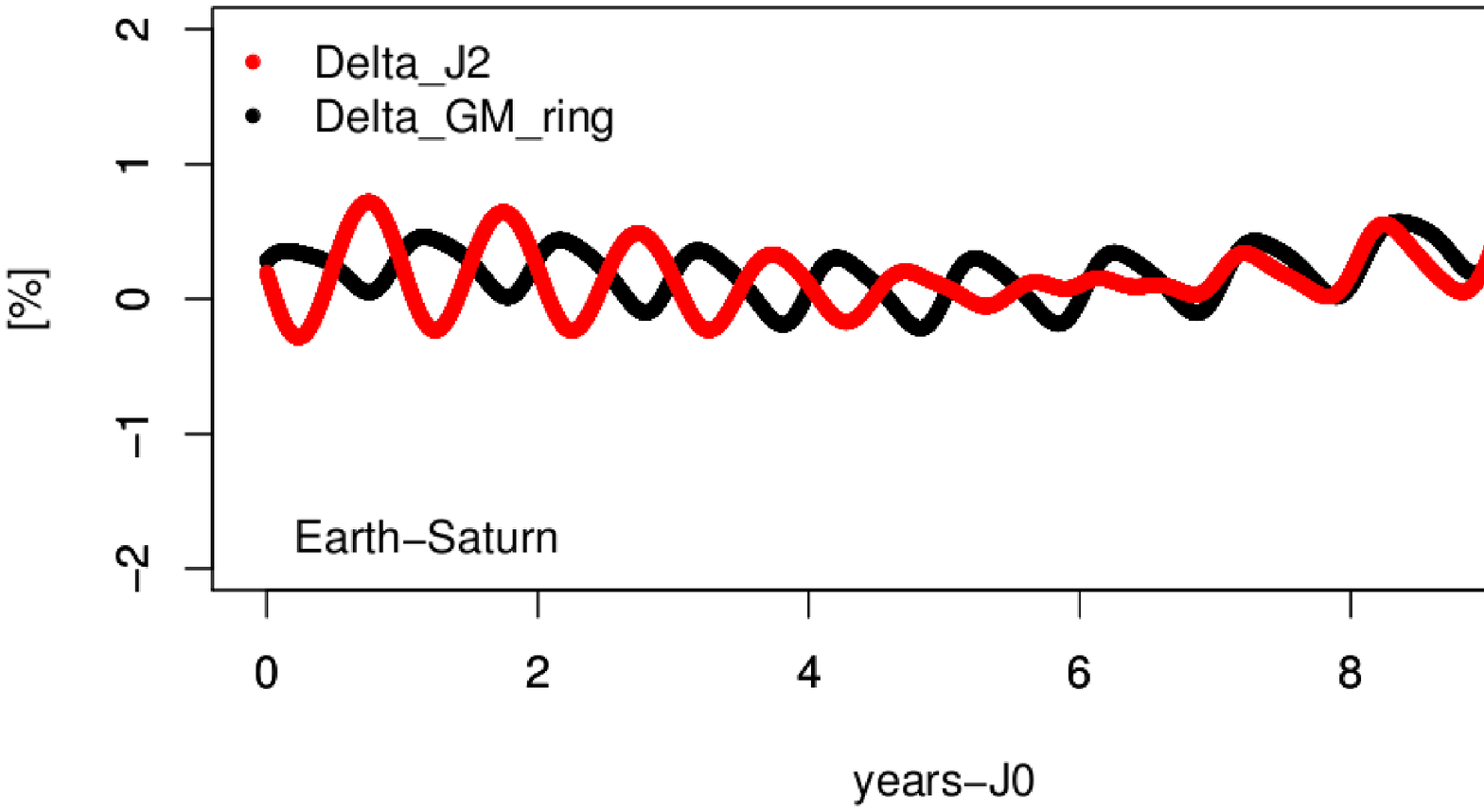}
\caption{Residuals obtained by comparisons between MGS/MO, VEX and Cassini range tracking data and ephemerides perturbed by a small change in sun J2 (12\%) and by a small change in the mass of the asteroid ring (17\%).} 
\label{impactmgsvexcass}
\end{center}
\end{figure*}

As illustrated on figure \ref{impactmgsvexcass}, there is a correlation between the effect on the geocentric distance of the modeling of the ring as done in INPOP08 in one hand and the effect of the sun oblatness in the other hand. 
Indeed, on these plots, one may see how a small change in the value of the sun J2 (12\%) induces after the refit of the planet initial conditions a periodic effect very similar in amplitude and frequency on Mercury, Mars and Venus distances to the Earth than a change in the mass of the asteroid ring (17\%).
Besides, the Saturn-Earth distances are not affected in the same way. As the ring does not affect the outer planets as it does on the inner planet, it becomes possible to decorrelate the signal induced by a small change in J2 from the one procuced by the ring when one can note an inversion of the GM signal for example in 2006. 
This result is consistent with the analytical study done by Iorio (2007) concluding also to the importance of taking into account the effects of asteroids on planetary orbit during relativistic tests.

It stress also the crucial importance of having a modeling of the asteroid perturbations as a fixed ring characterized independantly from the fit of planetary ephemerides.

We limit then a surestimation of the value of the sun J2 melting in this value some effects induced  by the asteroids.

Two different but complementary analysis and determination of PPN $\beta$ and Sun J2 are presented in the next sections with a fixed model of asteroid perturbations (same values of asteroid and ring masses and of densities as INPOP08).

\subsection{Estimations by least squares}

The first approach is based on the classic least square estimation of parameters during the fit of planet equations of motion to observations.
To check numerically the simplied assertion made in introdcution (section \ref{intro}), we estimate here what is the impact of each datasets in the determination of J2 and $\beta$: several adjustements of the initial condition of planets and the parameters J2 and $\beta$ are made using different sets of observations.
This leads to 32 adjustments based on INPOP08.
For each fit, changes were made in the selection of Mars and Venus data in order to estimate the impact of each important set of observations in the fit of the Sun J2 and PPN $\beta$. 
We look at the variations in the estimation errors of the 2 parameters and we use the 1-$\sigma$ given by the least squares as indicator of this uncertainty. With this method, we are then able to quantify the influence of each data sets on the determination of the pair $(\beta, J_2)$ as well as the stability of the determinations of the parameters. Indeed these variations in the error's estimation of the pair $(\beta, J_2)$ are a relevant indicator of the uncertainty of the fit of $\beta$ and J2.

 To take into account the correlation between $J_2$ and $\beta$, we use two modes of adjustments: in the {\it mode1}, $\beta$ or $J_2$ are fitted alone with the initial conditions of planets; in the {\it mode2}, both parameters are fitted simultaneously with the initial conditions of planets.

The results are summarized in table \ref{J2beta2}. One can first notice that the determinations of sun $J_2$ and $\beta$ made separately (i. e. mode1) give better $\sigma$ than fits including simultaneous $(\beta, J_2)$ determination (mode2). This is obviously consistent with the expected result relative to the determination of correlated parameters. 
The best results for a correlated determination of $J_2$ and $\beta$ (mode2) are then obtained when only the most accurate observations of Mars (MGS/MO, MEX and Vking) and Venus (VEX) are used simultaneously. 

Moreover, we note that the combined use of Venus ranging data and the complete data set for Mars do not really improve the separated determination (mode1) of $\beta$ and $J_2$, mainly due to the low accuracy of these observations, but a contrario it gives better correlated estimations (mode2). This is also consistent with the fact that fitting over observations from two different planets helps to decorrelate safely $J_2$ and $\beta$.
Furthermore, the Viking data by prolonging the fit interval with observations of rather good accuracies allow a decreasing of the uncertainties of about 20 $\%$ for $J_2$ and about 10$\%$ for $\beta$. Finally, it appears that the VEX data improves the determinations in a significant way:  decreasing of 31 $\%$ of the least squares $\sigma$ of the $J_2$ estimation and 48 $\%$ for $\beta$. Less than 2 years of VEX data have a bigger influence than a large interval (more than 30 years) of accurate Mars observations.
This is especially relevant for the PPN parameter $\beta$ with an improvement of about 48$\%$ of the accuracy between a determination including only Mars data and another one with both Mars and VEX data. In the same time, the improvement induced by the addition of Viking data is about 20 $\%$ for the $J_2$ and 10$\%$ for the PPN parameter $\beta$.
These figures show the crucial role of the VEX data before the use of future data from the ongoing generation of Mercury orbiters.

\begin{table}
\caption{1-$\sigma$ least squares obtained for J2 and $\beta$ using several sets of observations.} 
 \begin{tabular}{l c l l  l c l l}
\hline
& mode & J2  & $(1-\beta) $ & & mode & J2 & $(1-\beta) $\\
& & $\times 10^{7}$ & $\times 10^{3}$ & & & $\times 10^{7}$ & $\times 10^{3}$ \\
\hline
Modern Mars & 1 & 0.181 & &Impact of VEX & 1 & 0.144 & \\
MEX + MGS/MO& 1 & & 0.042 & Mars + VEX& 1 & & 0.025 \\
& 2 & 0.367 & 0.085 & & 2 & 0.208 & 0.037 \\
\\
Impact of Vkg & 1 & 0.161 & & \small{Impact of old Venus} & 1 & 0.188 & \\
MEX + MGS/MO + Vkg & 1 & & 0.040 & Mars + old Venus & 1 & & 0.040 \\
 = Mars & 2 & 0.302& 0.076 & &  2 & 0.283 & 0.060 \\
\hline
 \end{tabular}
\label{J2beta2}
\end{table}

\subsection{Incremental method and sensitivity estimation}
\label{incJ2}

An original strategy to study the sensitivity of the planetary ephemerides to J2 and PPN $\beta$ 
is to estimate how does differ from INPOP08 an ephemerides built using different values for J2 and PPN $\beta$ and fitted on the same set of observations as INPOP08. 
Such differences give an indication on how observations are sensitive to these parameters and with what accuracy can we estimate a parameter such as $\beta$.

To test such sensitivity, we focus our attention on the postfit residuals of the most accurate dataset used in INPOP08 adjustement: the Mercury direct range because of its sensitivity to general relativity and to the sun J2, VEX, MEX and MGS/MO data because of their high accuracy and simulated $S/N$ presented on table \ref{GRJ2} and the Jupiter Galileo data and Saturn Cassini normal point. These 2 latest data sets are selected because they induce a global improvement of the planetary ephemerides in its all and especially of the Earth orbit. 

\begin{figure*}
\begin{center}
\includegraphics[width=5.9cm]{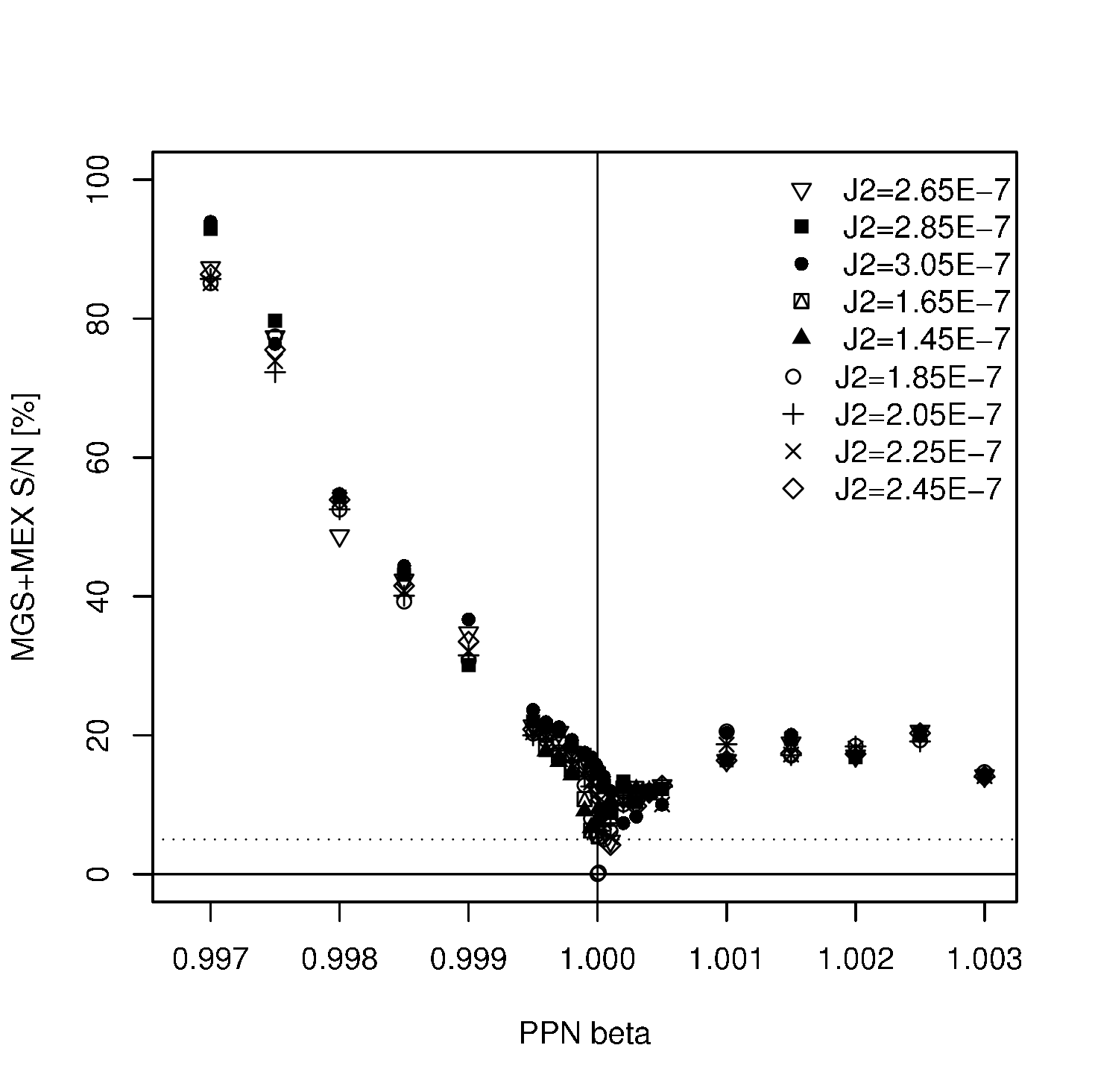}\includegraphics[width=5.9cm]{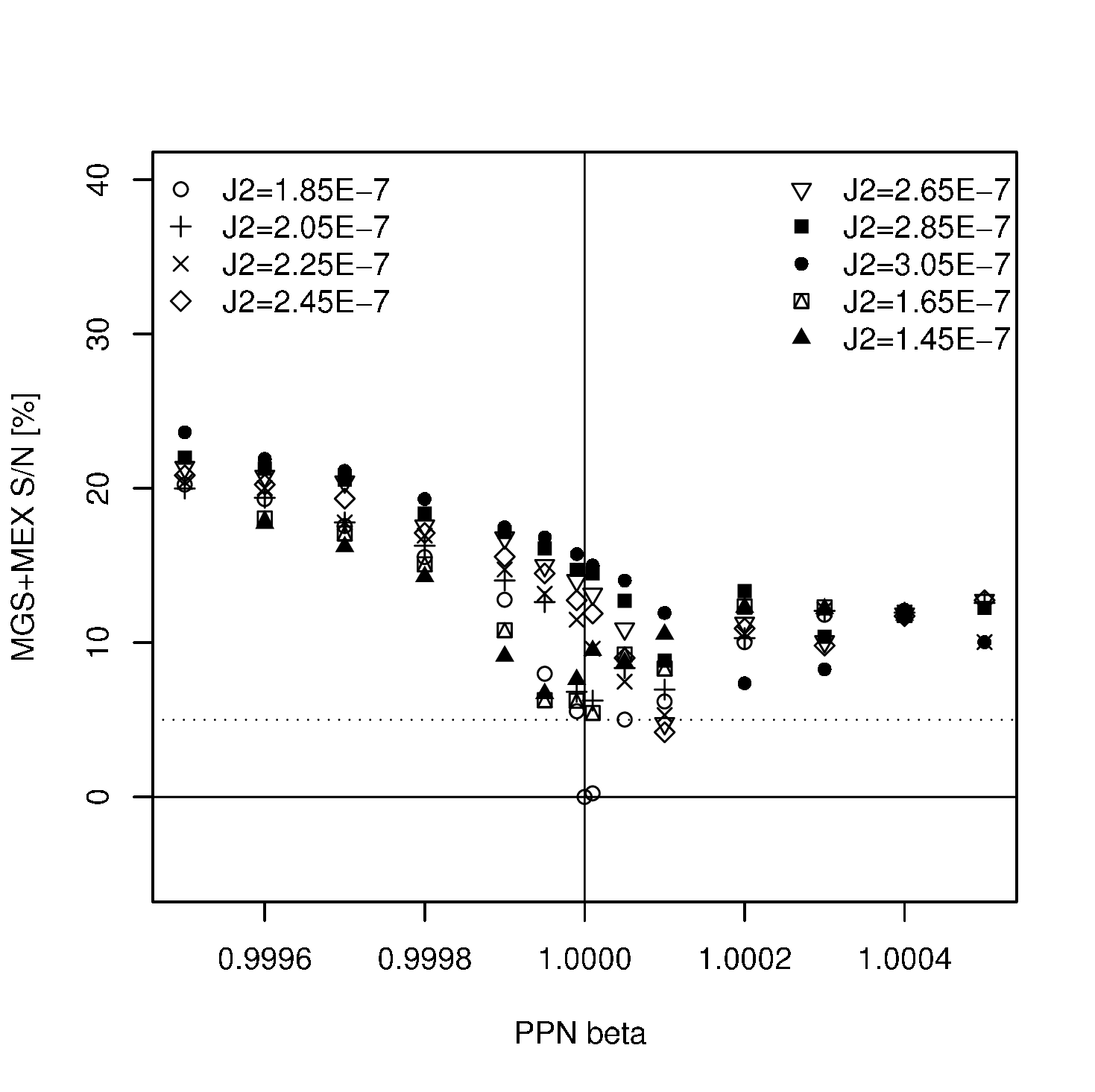}\includegraphics[width=5.9cm]{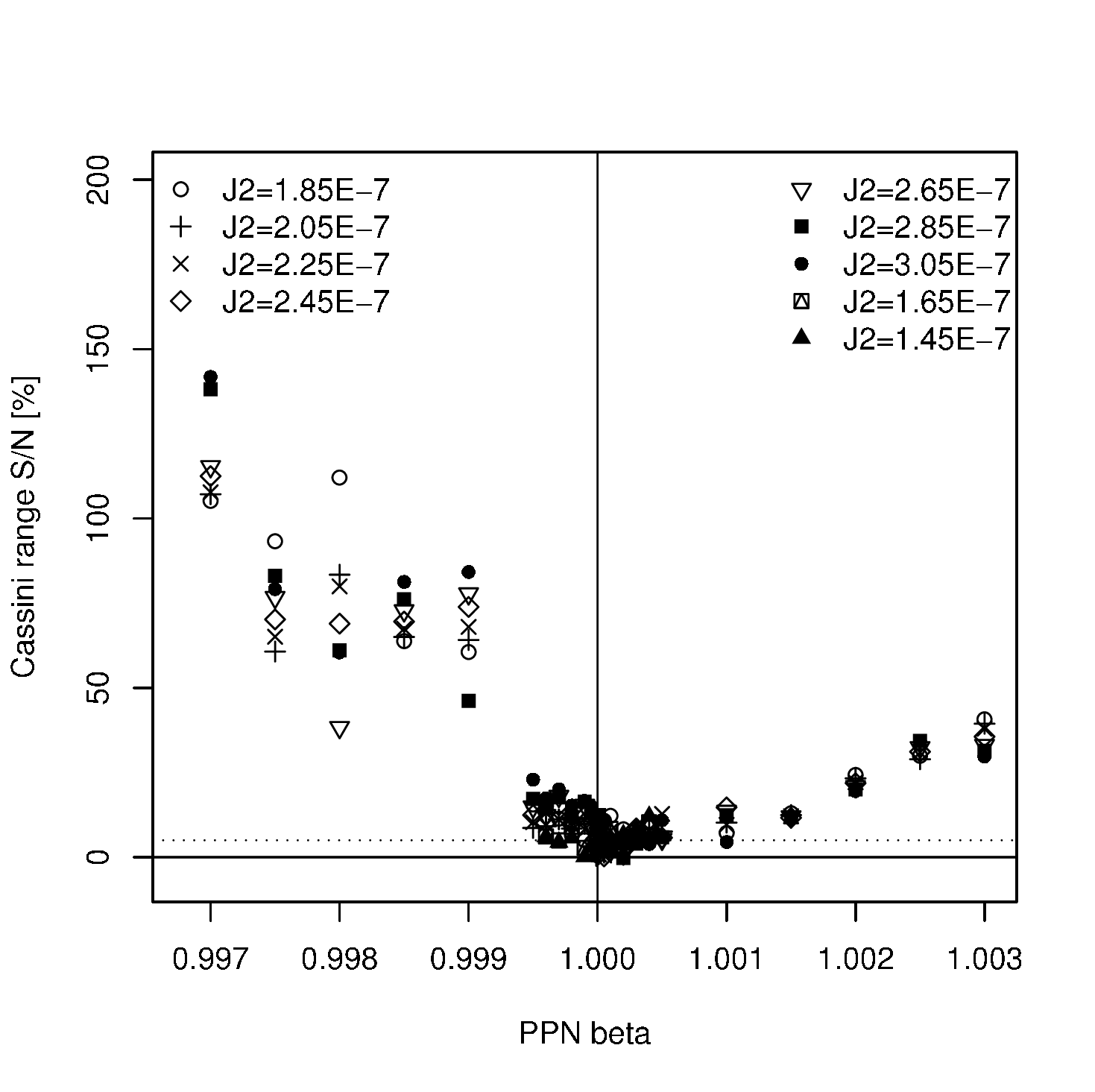}
\includegraphics[width=5.9cm]{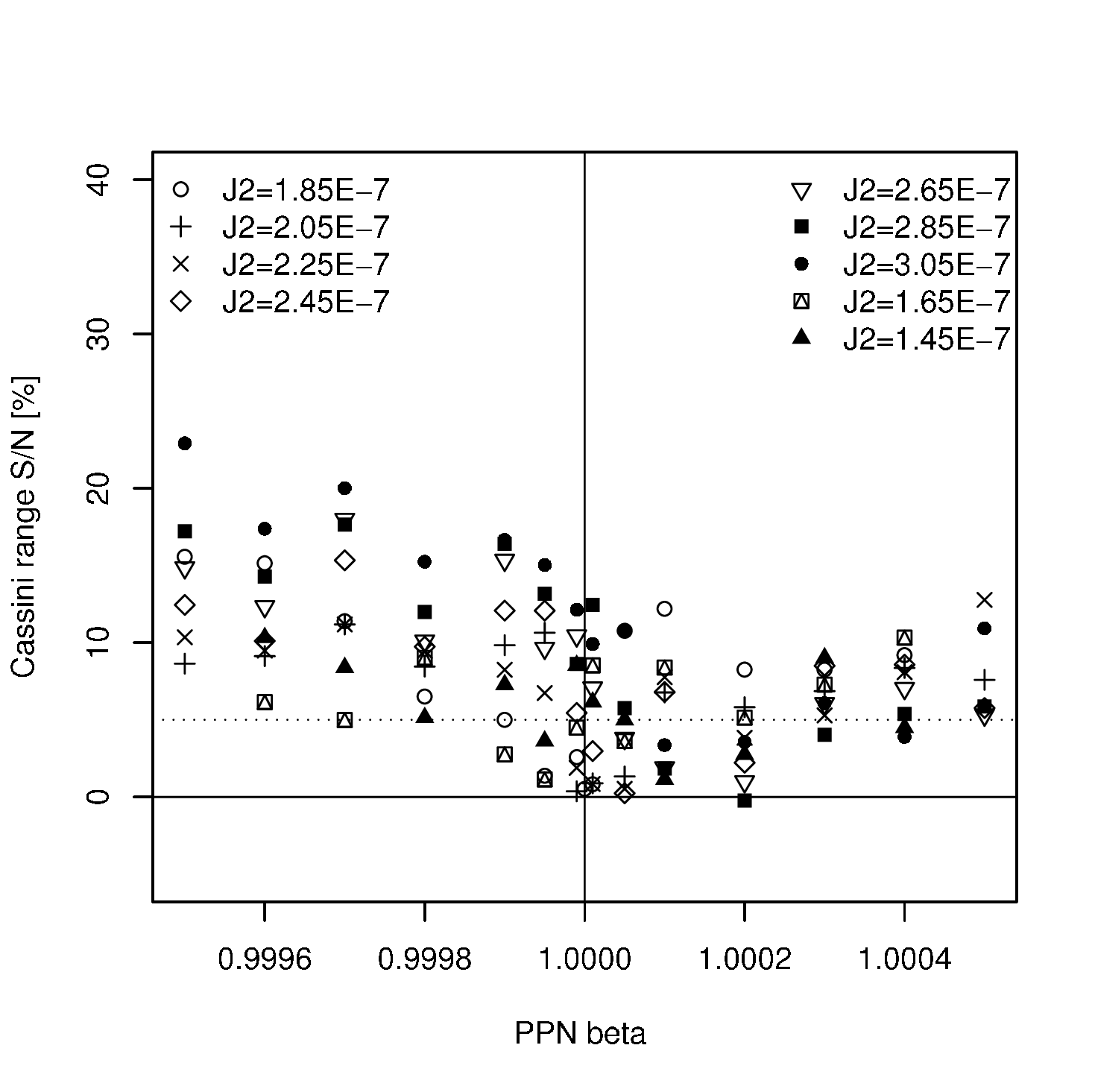}\includegraphics[width=5.9cm]{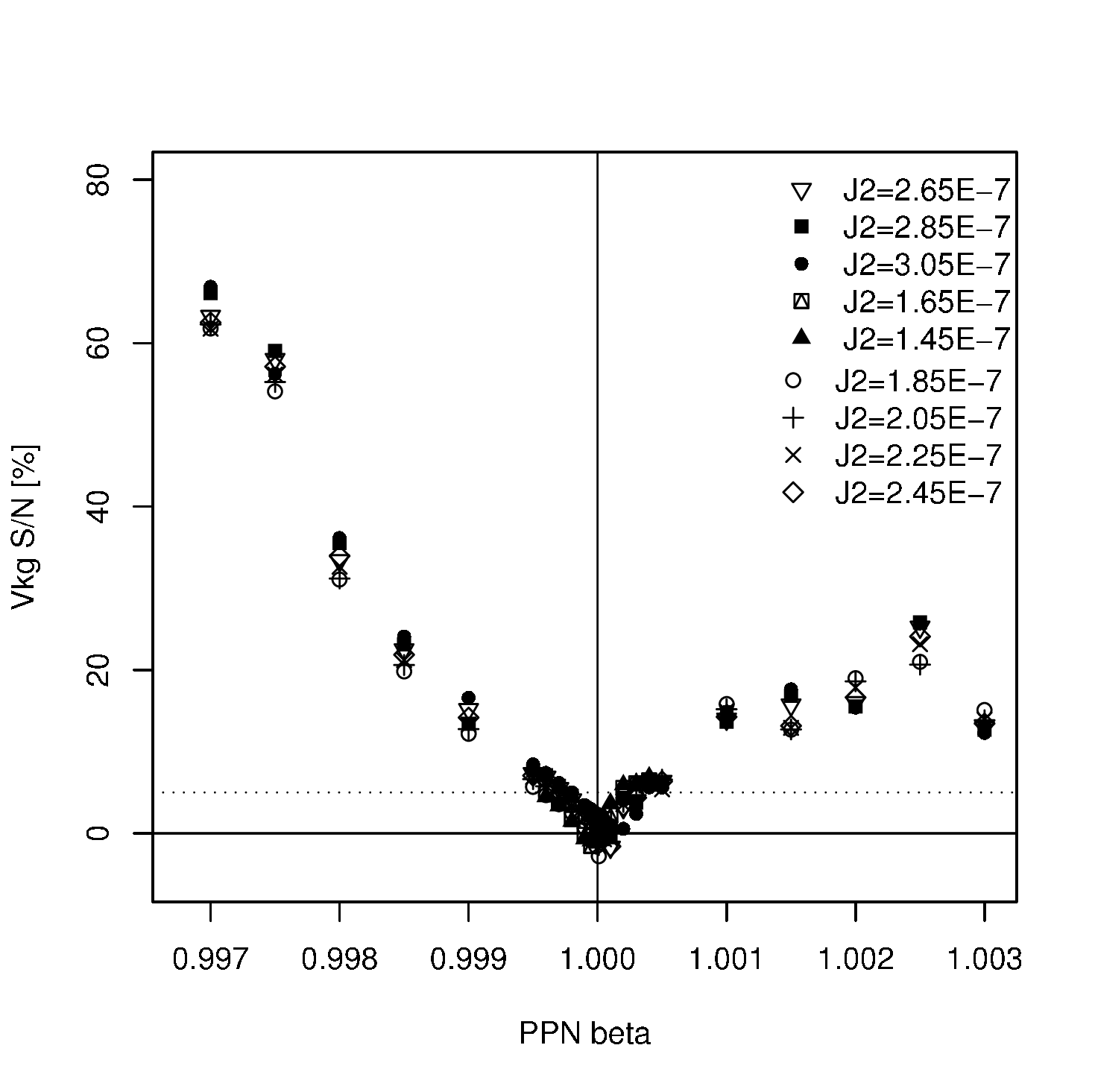}\includegraphics[width=5.9cm]{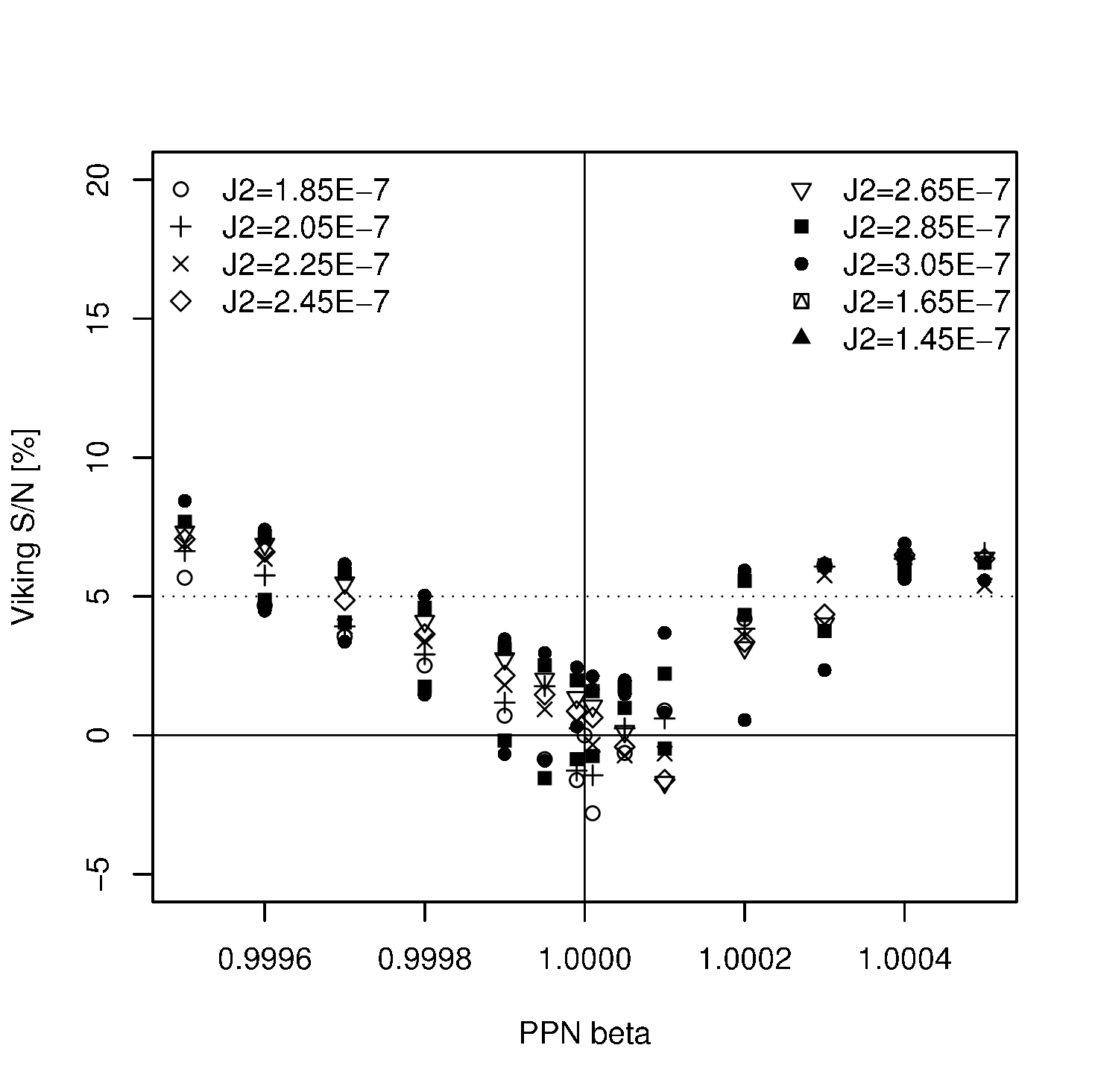}
\includegraphics[width=5.9cm]{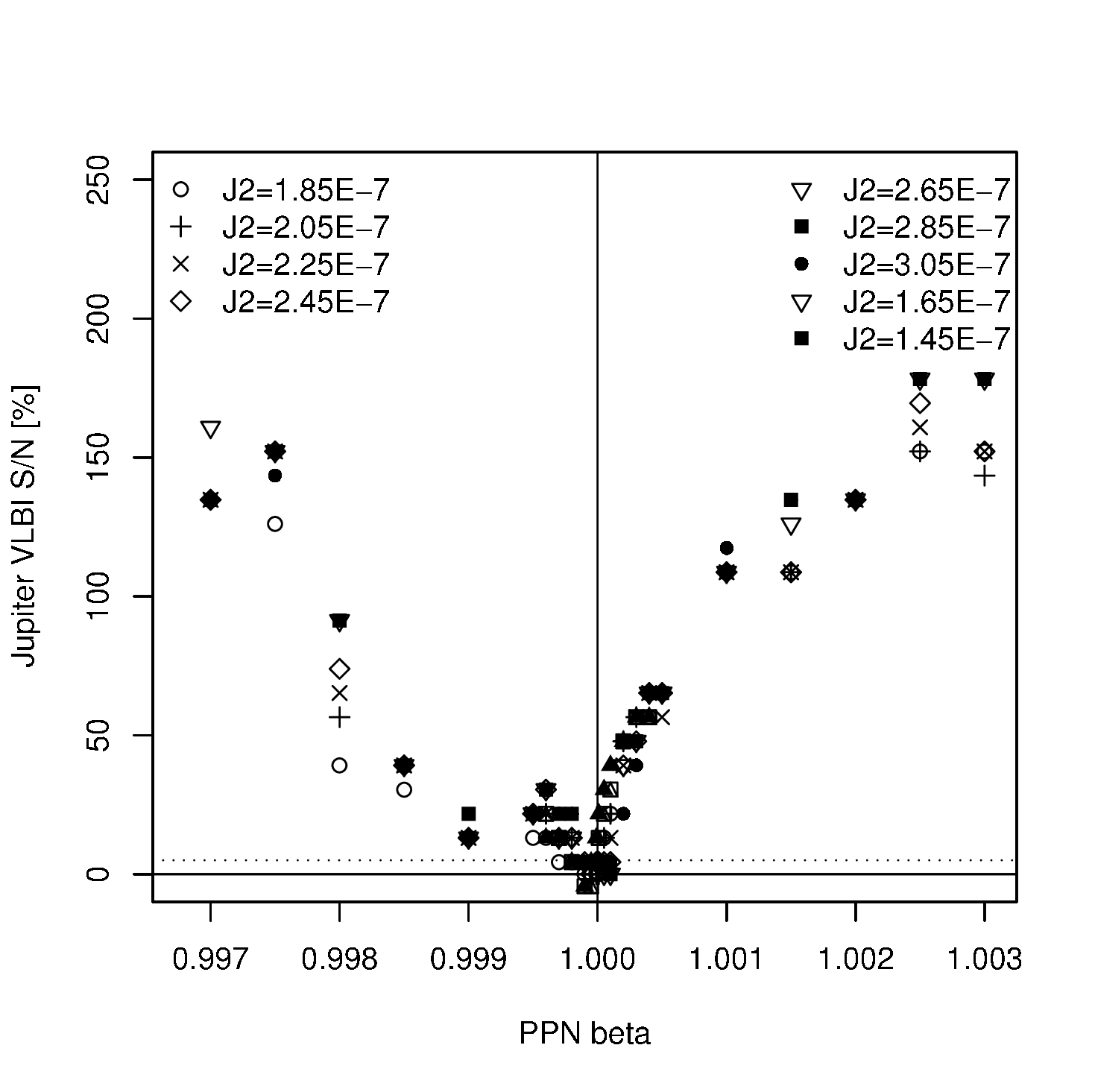}\includegraphics[width=5.9cm]{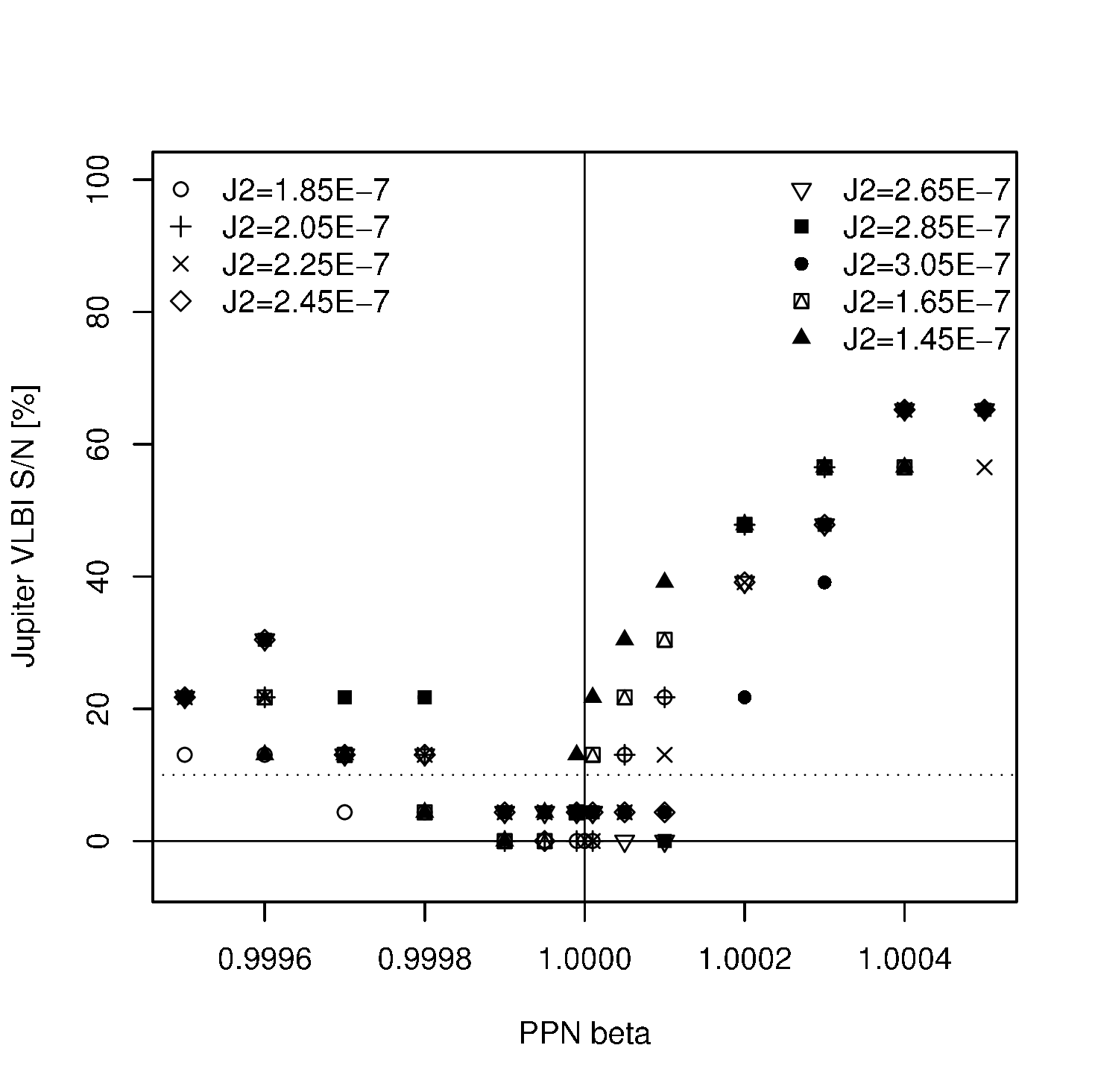}\includegraphics[width=5.9cm]{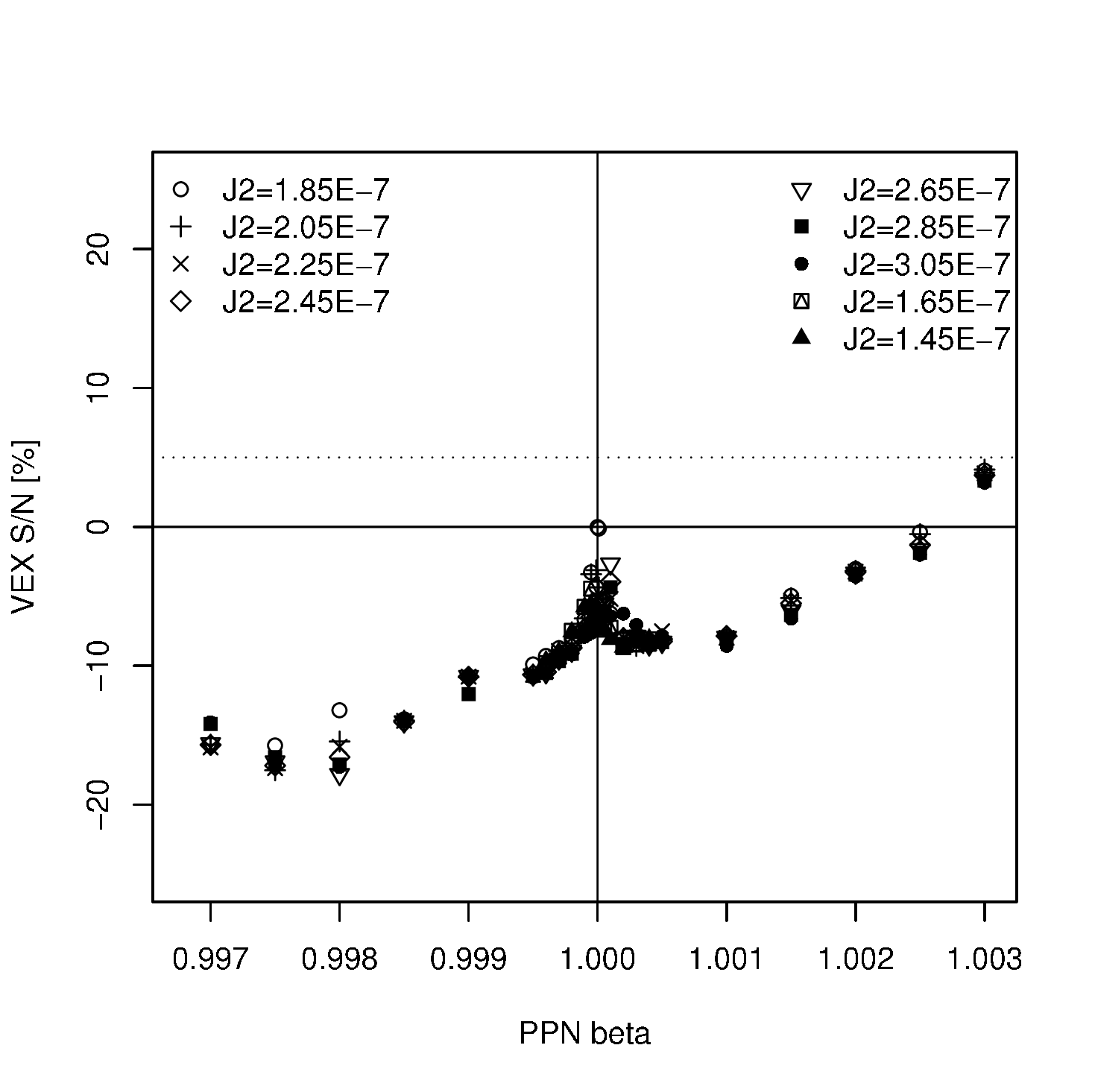}
\includegraphics[width=5.9cm]{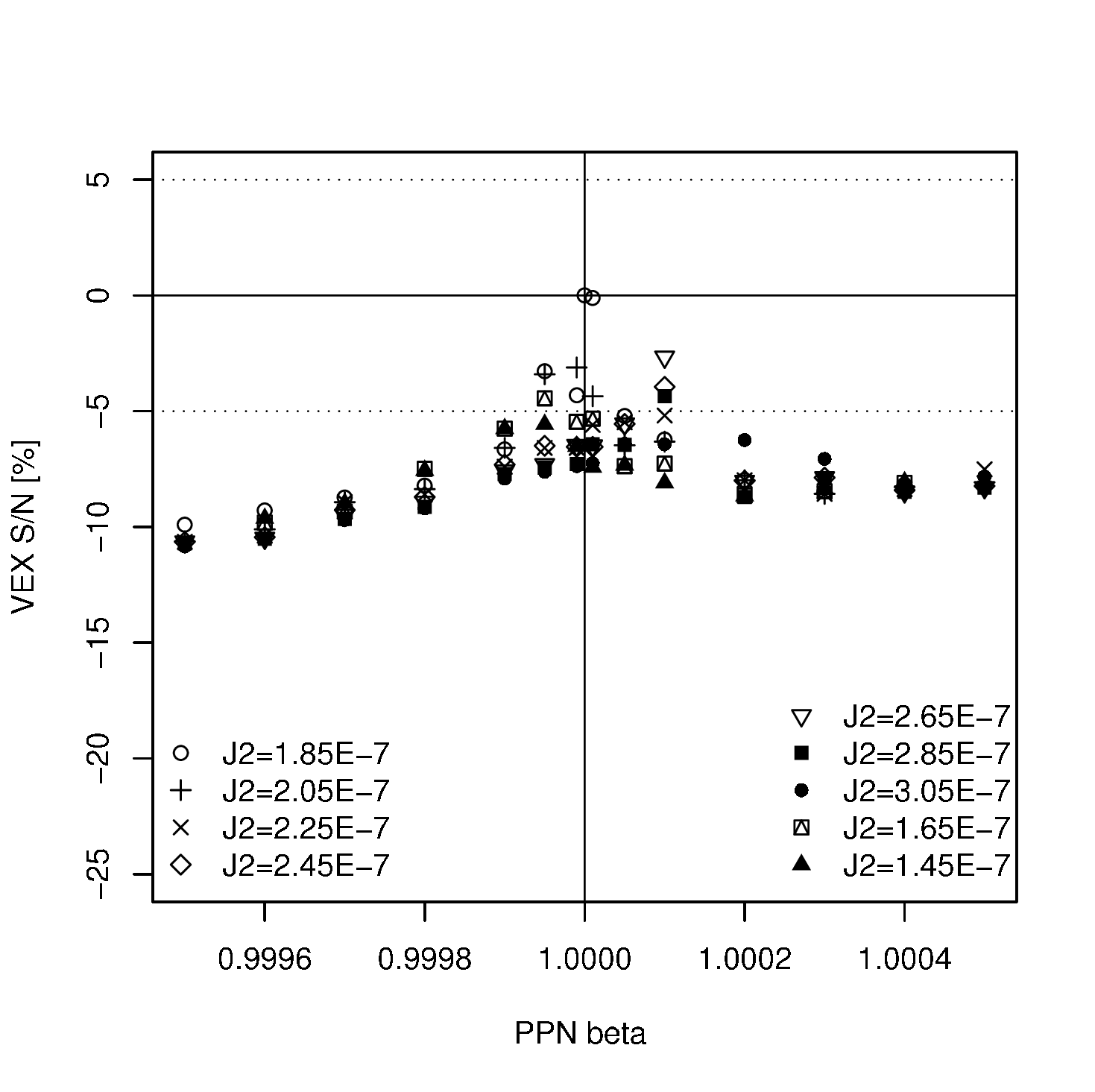}\includegraphics[width=5.9cm]{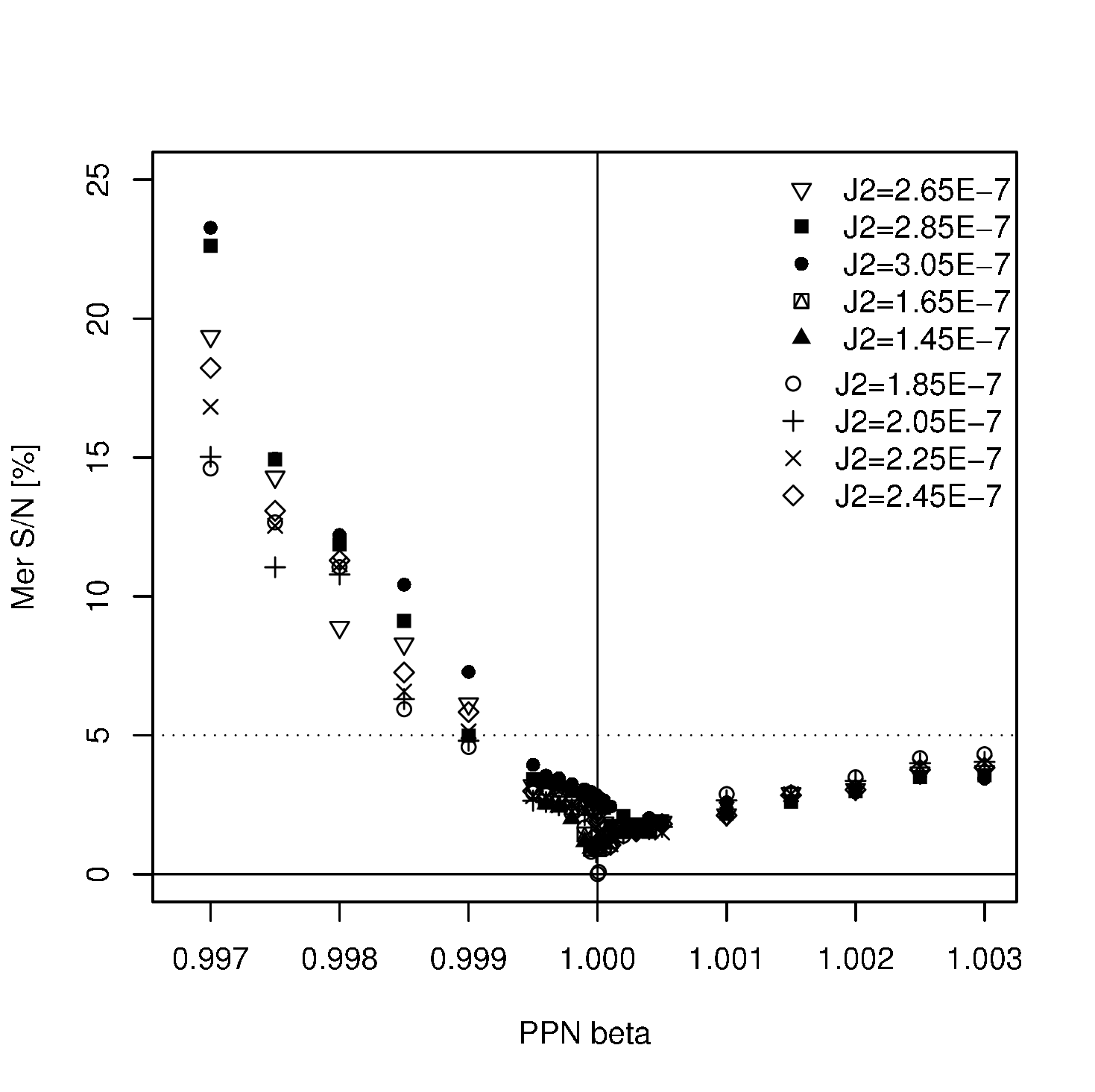}\includegraphics[width=5.9cm]{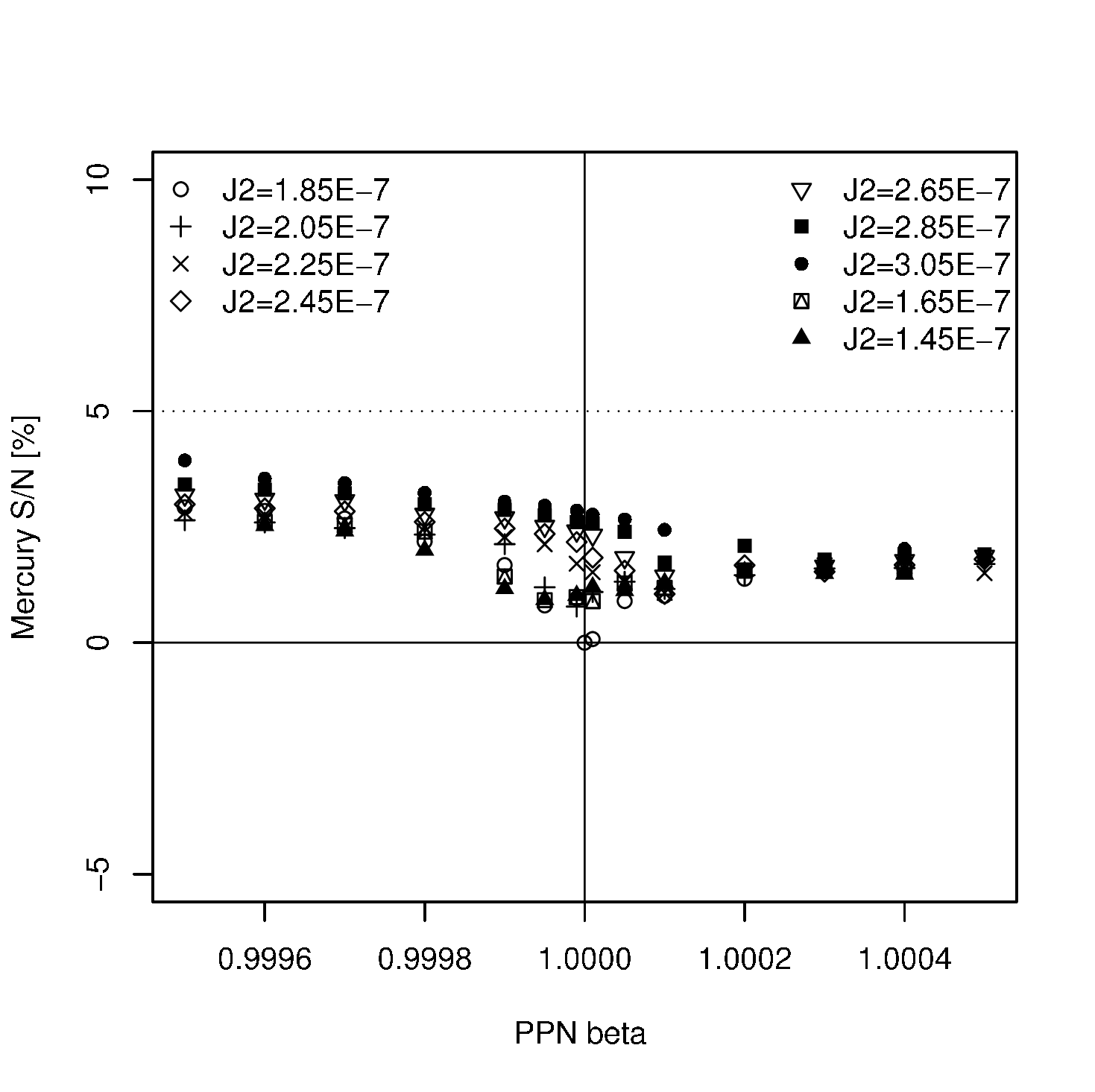}
\caption{Residuals obtained by comparisons between observations and ephemerides estimated with different values of PPN $\beta$ (values given on x-axis of each subframes) and different values of Sun $J_2$.  } 
\label{mapJ2beta1}
\end{center}
\end{figure*}

To estimate the sensitivity of these 7 most accurate sets of data used in INPOP08 adjustment to the variations of values of J2 and PPN $\beta$, we have estimated and plotted the ratio $S/N$ defined as:
$$
S/N = \frac{\sigma_{i,j}-\sigma_{0,0}}{\sigma_{0,0}}
$$
where $\sigma_{i,j}$ is the 1-sigma dispersion of the postfit residuals of an ephemerides based on INPOP08 but with values of J2 and PPN $\beta$ different from the ones used in INPOP08 (which are $\beta=1.0$ and $J2 = 1.82 \times 10^{-7}$) and fitted on all the INPOP08 data sets and $\sigma_{0,0}$ is the 1-sigma dispersion of the postfit INPOP08 residuals.
We have used 9 values of J2 varying from $1.45 \times 10^{-7}$ to $3.05 \times 10^{-7}$ with a 0.2 step and 24 values of PPN $\beta$, building then 192 different ephemerides.
The 24 values of $\beta$ are distributed over 2 windows: a global one based on 12 values of $\beta$ varying from 0.997 to 1.003 with a 0.0005 step (window 1) and from 0.9996 to 1.0004 with a step of 0.0001 (window 2).
Results presented as the $S/N$ pourcentage, are plotted on figures \ref{mapJ2beta1} and \ref{mapJ2beta2}.

As one can see on figures \ref{mapJ2beta1}, the impact of the PPN $\beta$ is not symmetric to $\beta=1$. 
On figure \ref{mapJ2beta1}, one notices also the direct correlation between the $S/N$ obtained with MGS/MO and MEX data and the one obtained for VEX.

One may see on the figure \ref{mapJ2beta1} that the $S/N$ of the Jupiter and Saturn data sets are sensitive to changes in J2 and PPN $\beta$. 
The sensitivity of these datasets are not crucial for the analysis but they reflect the impact of the use of such observations in the improvement of the Earth orbit and then the sensitivity of the Earth orbit to the gravity testing.
On table \ref{betamaxmin}, we have gathered minimum and maximum values of PPN $\beta$ defining the sensitivity interval of the different datasets. 
The sensitivity interval is the interval of PPN $\beta$ for which the $S/N$ remains below 5$\%$. 
Values of PPN $\beta$ greater than the maximum value given in table \ref{betamaxmin} or smaller than the minimum value cannot be seen as realistic in comparison to modern observations.
By considering the figures \ref{mapJ2beta1} and \ref{mapJ2beta2} and the table \ref{betamaxmin} it appears that the MGS/MO and MEX data provide the most narrow interval of sensitivity with $0.99995 < \beta < 1.0002$. 
This interval is in agreement with the latest determinations done by (Williams et al., 2009), (Fienga et al 2008) and (Pitjeva 2006). 

\begin{table}
\caption{$\beta$ interval in which the residuals stay below the 5$\%$ limit (upper table) and comparisons between several other estimations found in the literature (bottom table). In the case of (Williams et al. 2009), the value of $\beta$ given here is estimated for $\gamma=1$.} 
 \begin{tabular}{l l l | l l l | l l l }
\hline
Data & $\beta$ min & $\beta$ max & Data & $\beta$ min & $\beta$ max & Data & $\beta$ min & $\beta$ max\\
\hline
MGS/MO+MEX & 0.99995 & 1.0002 & Jupiter VLBI & 0.9996 & 1.0002 & Viking & 0.9995 & 1.0002\\
VEX & 0.99990 & 1.0002 & Saturn Cassini range & 0.9998 & 1.0005 & Mercury & 0.9985 & 1.005\\
\hline
\end{tabular}
\label{betamaxmin}
\end{table}

\section{Secular advances of planetary perihelia}
\label{Pi}

We are interested here by evaluating if the observations used to fit INPOP08 would be sensitve to supplementary precession of the planet orbits.
Such precessions would have be seen as supplementary secular advance of the orbit perihelion unexplained by general relativity.  
To estimate the sensitivity of the modern tracking data, 
we first fix J2=$1.8 \times 10^{-7}$, $\beta=1$ and $\gamma=1$. 
By fixing the value of the sun J2, we then isolated the impact of the secular advance of the perihelion for one given value of J2.

\begin{figure*}
 \begin{center}
\includegraphics[width=15cm]{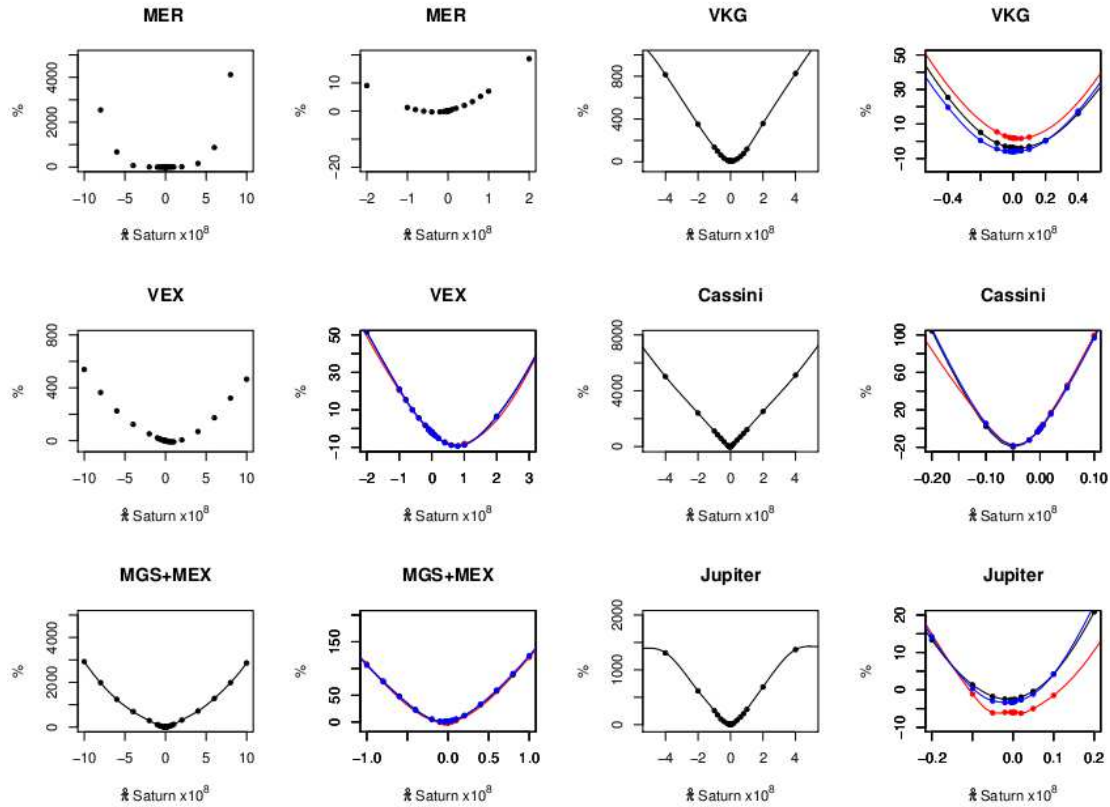}
\end{center}
\caption{Residuals obtained by comparisons between observations and ephemerides estimated with different values of $\dot{\pi}$}
  \label{pixfig}
 \end{figure*}

For each different values of $\dot{\pi}$, initial conditions of planets are fitted to the INPOP08 observations and we compare the postfit residuals to the INPOP08 ones.
We focused our study for the same sets of observations as for the J2,$\beta$ study. 
As one can on figure \ref{pixfig}, the behaviour of the obtained S/N  (as defined in section \ref{incJ2}) is symmetrical to a minimum value, this minimal value being centered around $\dot{\pi}=0$ or not. This symmetry explains why in table \ref{pimaxmin} we gives interval of $\dot{\pi}$ for which the minimum of S/N is obtained. 
One can then compared these values to those published by (Pitjeva 2009). 
For all planets, except Saturn, the values of $\dot{\pi}$ minimizing the residuals are not significantly different from zero. 
One can note that the best constraint on the Earth orbit is given by the Jupiter VLBI datasets which gives the narrowest interval of $\dot{\pi}$.
For Saturn, an offset in the minimum of the S/N is obtained for the Cassini tracking data sets ($-10 \pm 8$) and the VEX datasets ($200 \pm 160$). 
These estimations lead to two statistically significant and different determinations of a supplementary precession of the Saturn orbit. 
By comparisons, (Pitjeva 2009) value is very close to the one we obtain in considering only the S/N induced on the Cassini observations. 
This result shows how important is the description of the method used to evaluate such quantities.

To test the stability of the estimations and as it is well-known that the asteroids induce a global precession of the inner planets perihelia, we operate the same computations with small changes in the mass of the ring (20\%) and in the Sun J2 (5\%) values. The obtained variations of the S/N are plotted on figure \ref{pixfig} where the red curves are the results obtained with the change in the mass of the ring and the blue curves are the ones deduced from the J2. Some changes are noticeable for Viking and Jupiter, however, for Cassini and VEX, the minimum are stable. 

The investigation about  a statistically significant advance in the Saturn perihelion has to be continued in using more Cassini and VEX data. 
Indeed, a prolongation of the interval of time covered by these two datasets will improve the accuracy of the estimations.

\begin{table}
\caption{$\dot{\pi}$ intervals minimizing postfit residuals.} 
 \begin{tabular}{l l l l l l l l l}
\hline
Data & $\dot{\pi}$ Mer &  $\dot{\pi}$ Ven & $\dot{\pi}$ EMB & $\dot{\pi}$ Mars & $\dot{\pi}$ Jup &  $\dot{\pi}$ Saturn & $\dot{\pi}$ Ura & $\dot{\pi}$ Nep\\
mas$/$cy& & & & && &$\times 10^{-4}$ & $\times 10^{-4}$\\
\hline
Mercury & -10 $\pm$ 30 & 30 $\pm$ 130 & 0 $\pm$ 40& $>$ 2000 & $>$ 2000&  0 $\pm$ 200  & $> 20$  & $> 20$\\
VEX & 0 $\pm$ 200 & 18 $\pm$ 22 & 0 $\pm$ 4& 0 $\pm$ 1.4 & 0 $\pm$ 200  & 200 $\pm$ 160   & 0 $\pm$ 2  &  $> 20$\\
\tiny{MGS/MO+MEX} & 0 $\pm$ 200 & -24 $\pm$ 34 & -0.4 $\pm$ 0.8 & 0.4 $\pm$ 0.6 & -20 $\pm$ 180  & 0 $\pm$ 60  & 0 $\pm$ 2 & 0 $\pm$ 10\\
Viking & 0 $\pm$ 200 & -24 $\pm$ 34 &  0 $\pm$ 0.2 & 0 $\pm$ 0.2 & -200 $\pm$ 200  & 0 $\pm$ 10  & (4 $\pm$ 4) & 0 $\pm$ 10\\
\small{Jupiter VLBI} & 0 $\pm$ 400 & -4 $\pm$ 6 & 0 $\pm$ 0.016& 0 $\pm$ 0.6 & 142 $\pm$ 156   & 0 $\pm$ 10    & 0 $\pm$ 2 & 0 $\pm$ 2 \\
\small{Saturn range} & $>$ 2000 & 0 $\pm$ 10 & 0.1 $\pm$ 0.1& 0 $\pm$ 0.2 & 0 $\pm$ 400 &  -10 $\pm$ 8 & 0 $\pm$ 2  & 0 $\pm$ 2 \\
 \small{Cassini} & & & & &&&&\\
\hline
\hline
\\
Pitjeva 2009  & -3.6 $\pm$ 5 & -0.4 $\pm$ 0.5 & -0.2 $\pm$ 0.4 & 0.1 $\pm$ 0.5&  & -6 $\pm$ 2 &  &\\
\hline
\end{tabular}
\label{pimaxmin}
\end{table}

\section{Does the Pioneer anomaly impact the ephemerides ?}
\label{PA}

Since 2002 and the confirmation by several teams of the detection of acceleration anomalies in the tracking of several spacecrafts, three classes of possible explanations were given; first, the detected acceleration is not really an acceleration but is an avatar or a manifestation of a mis-modeling in the Doppler and ranging signals taped by navigation teams. Second, the anomaly is a mis-modeling in the orbit of the probe itself induced by a technical problem or misunderstandings of the spacecraft techniques. The third cause invocated is a generalization of the second by implying a mis-modeling in the dynamics of the probe but also of all objects in the solar system and beyond.
Thus, if the equivalence principal is followed, the equations of motion of the major planets of our solar system have also to be modified in the same manner as the spacecraft dynamical equations are.


We investigate this question by using the INPOP08 planetary ephemerides as a test bed for some hypothesis describing the pioneer anomalies.

\begin{figure*}
 \begin{center}
\includegraphics[width=18cm]{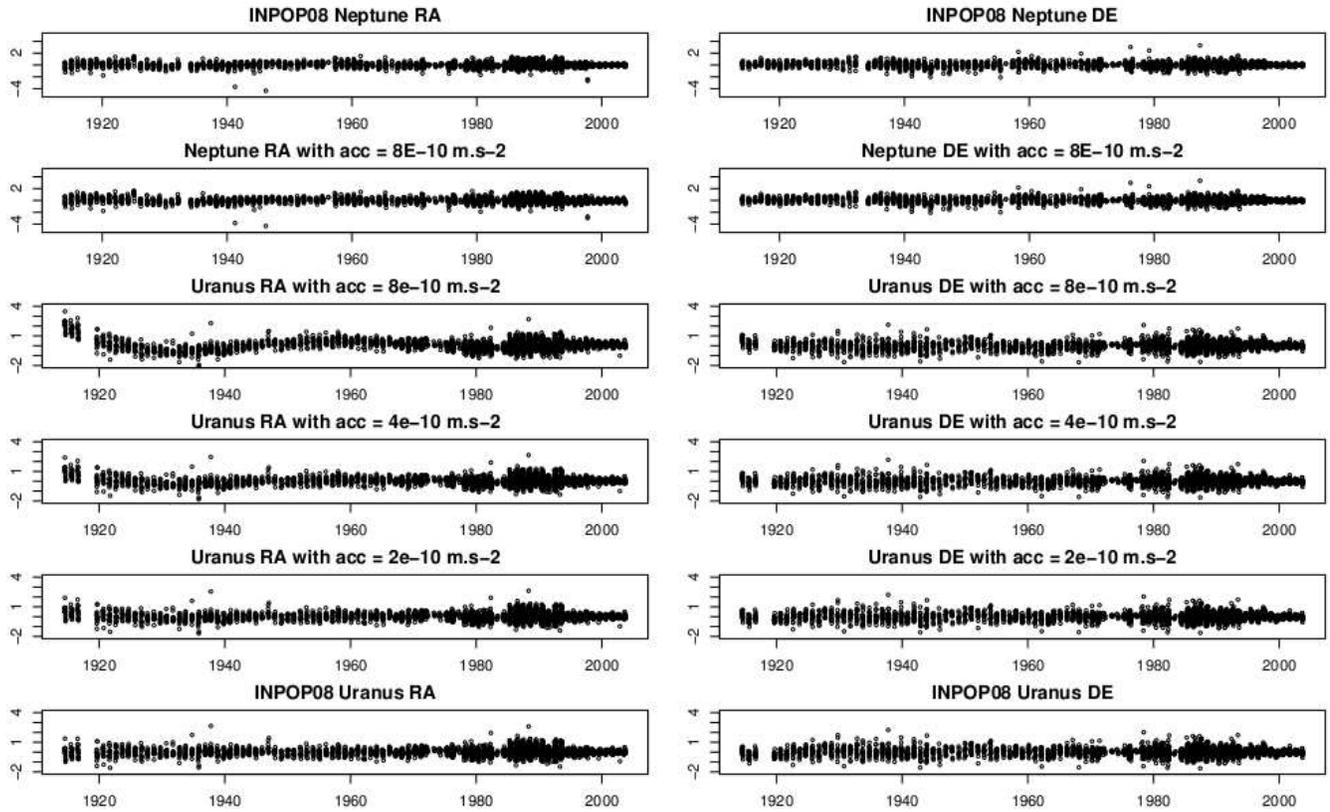}
\end{center}
\caption{Residuals in right ascention and declination of Neptune and Uranus  obtained with INPOP08 (solution of reference) and a fitted solution including PA of different magnitudes: from 8 to 2 $\times 10^{-10}$ $m.s^{-2}$. The x-axis are years and y-axis is in arcseconds.}
  \label{omcuranus}
 \end{figure*}


A classic description of the pioneer anomalies(PA)  is the apperearance of an constant acceleration of about $ 8.75 \times 10^{-10} m.s^{-2}$, Sun-oriented after 20 AU (Anderson et al. 2002a).
We then add this constant acceleration in the equations of motions of Uranus, Neptune and Pluto. This constitutes the modified ephemerides as noted in the following.

We have fitted the modified ephemerides to observations usually used to built INPOP08. 
Residuals obtained after the fit are plotted in Figure \ref{omcuranus}. 
The value of the acceleration was changed in a way to obtain a minimum value for which the effect induced by such acceleration becomes detectable in the residuals.
As it appears clearly in the residuals of Uranus right ascention, a constant acceleration of $8 \times 10^{-10} m.s^{-2}$ added to the classical Einstein-Hoffmann
equation of motions can not be missed even after the fit of the initial conditions of planets. 
A systematic effect remains especially after 1930. 
This effect cannot be absorbed by the fit or by the noise of the old Uranus observations made at that time. 
By changing the value of the acceleration, one sees that the acceleration must be at least 4 times smaller than the one commonly adopted to be 
absorbed by the residuals.
For Neptune and Pluto, the situation is different.
For these planets, the effect of a constant acceleration is absorbed by the fit, as one can see on figure \ref{omcuranus} with the postfit and prefit residuals of Neptune.

\section{Conclusions}

Concerning the determination of the PPN parameter $\beta$, an estimation of the planetary ephemerides sensitivity to this parameter is done follwing two methods. 
Our results show that a global fit is needed in order to decorrelate parameters such as PPN $\beta$, Sun J2 and the asteroid pertubations. 

We have tested possible detection of anomalous advance of perihela of planets. 
More investigations are needed for the analysis of the perihelion rate of Saturn and more observations of Cassini and VEX data are necessary.

Finally, the results obtained here for the Pioneer Anomaly induce that no constant acceleration larger than 1/4 the PA can affect the planets of our solar system. If it was so, it would have been detected sooner. In the frame of the equivalence principle, this means that no constant acceleration larger than 1/4 the PA can be realistic.

\end{document}